\documentclass[doublecol]{epl2} 

\usepackage[utf8]{inputenc}
\usepackage{amsmath,amssymb}
\usepackage{graphicx}
\usepackage[colorlinks=true,citecolor=blue]{hyperref}
\usepackage{units}
\usepackage{abbrevs}

\newabbrev\EMP{efficiency at maximum power (EMP)}[EMP]

\newcommand{\kBT}{k_\text{B}T}

\def\kB {k_\text{B}}

\DeclareMathOperator{\Det}{Det}
\DeclareMathOperator{\Li}{Li}

\title{Quantum Nernst engines}
\shorttitle{Quantum Nernst engines} 

\author{Björn Sothmann\inst{1} \and Rafael Sánchez\inst{2} \and Andrew N. Jordan\inst{3,4}}
\shortauthor{Bj\"orn Sothmann \etal}

\institute{                    
  \inst{1} Département de Physique Théorique, Université de Genève, CH-1211 Genève 4, Switzerland\\
  \inst{2} Instituto de Ciencia de Materiales de Madrid (ICMM-CSIC), Cantoblanco, E-28049 Madrid, Spain\\
  \inst{3} Department of Physics and Astronomy, University of Rochester, Rochester, New York 14627, USA\\
  \inst{4} Institute for Quantum Studies, Chapman University, Orange, California 92866, USA
}

\abstract{
We theoretically propose Nernst engines based on quantum Hall edge states. We identify a setup that exhibits an extreme asymmetry between the off-diagonal Onsager coefficients for heat and charge transport. In terms of thermodynamic efficiency, this engine outperforms a recently proposed classical Nernst engine. A second setup using an antidot is found to be more efficient as energy filtering becomes less strong.
}

\begin{document}

\maketitle

\section{\label{sec:Intro}Introduction}
Nanoscale thermoelectrics has recently received a lot of interest. On the one hand, this was driven by  fundamental questions that aimed, e.g., to understand the thermopower of basic mesoscopic devices such as quantum point contacts~\cite{streda_quantised_1989,proetto_thermopower_1991,molenkamp_quantum_1990} or quantum dots.~\cite{beenakker_theory_1992,staring_coulomb-blockade_1993,dzurak_thermoelectric_1997,godijn_thermopower_1999,scheibner_thermopower_2005,scheibner_sequential_2007,svensson_lineshape_2012,svensson_nonlinear_2013,thierschmann_diffusion_2013} On the other hand, nano heat engines are also promising candidates for energy harvesting applications~\cite{sanchez_optimal_2011,sothmann_rectification_2012,sothmann_magnon-driven_2012,jordan_powerful_2013,sothmann_powerful_2013,jiang_three-terminal_2013}. Three-terminal harvesters benefit from the possibility to separate charge and heat flows. In a four-terminal configuration, this can be taken to the extreme: two different pairs of terminals serve for the injection of charge and heat currents, respectively.

A central question that is relevant both from a fundamental as well as from an applied point of view concerns the limits on the efficiency of a heat engine operating between a hot and a cold reservoir with temperatures $T_1$ and $T_2$, respectively.

A first answer to this question was obtained by Carnot who showed that thermodynamics dictates the efficiency to be smaller than the Carnot efficiency $\eta_C=1-T_2/T_1$. By now, several theoretical works demonstrated the possibility of reaching $\eta_C$ in nanoscale heat engines~\cite{sanchez_optimal_2011,sothmann_magnon-driven_2012,ruokola_theory_2012,brunner_virtual_2012,jordan_powerful_2013,bergenfeldt_hybrid_2014,kennes_efficiency_2013}. However, this limit can only be reached for a heat engine that operates reversibly and, therefore, does not generate any output power. 

For applications it is therefore more relevant to analyze the \EMP $\eta_\text{maxP}$. Using the framework of linear irreversible thermodynamics it was shown that a system with time-reversal symmetry satisfies $\eta_\text{maxP}= (\eta_C/2)[ZT/(1+ZT)]$ where the figure of merit satisfies $ZT\geq 0$~\cite{van_den_broeck_thermodynamic_2005}. $ZT$ diverges in the tight-coupling limit where heat and charge currents are proportional to each other and $\eta_\text{maxP}$ takes its maximal value $\eta_C/2$.
While the above limit is completely universal, this is no longer true in the nonlinear regime where $\eta_\text{maxP}$ depends on the symmetry of the heat engine~\cite{esposito_universality_2009}.

For systems that break time-reversal symmetry, the \EMP not only depends on the figure of merit but also on the asymmetry of the off-diagonal Onsager coefficients~\cite{benenti_thermodynamic_2011}. For a finite asymmetry, the \EMP can overcome the bounds imposed by thermodynamics in the presence of time-reversal symmetry and even reach Carnot efficiency~\cite{benenti_thermodynamic_2011}. 
However, in multi-terminal setups, current conservation imposes additional constraints that lead to an \EMP smaller than $\eta_C$ but can be larger than the bound $\eta_C/2$ valid in the presence of time-reversal symmetry~\cite{brandner_strong_2013,brandner_multi-terminal_2013}.

A particular realization of such multi-terminal setups breaking time-reversal symmetry is given by a Nernst engine where a magnetic field is applied perpendicular to a two-dimensional conductor. A temperature bias along the sample will then give rise to a charge current perpendicular to both the temperature gradient and the magnetic field. Recently, such a Nernst engine based on transport of classical particles in a four-terminal setup has been analyzed in Ref.~\cite{stark_classical_2014}. This classical Nernst engine has off-diagonal Onsager coefficients of equal magnitude and opposite sign. It was shown to saturate the resulting bounds on efficiency and EMP. Experimentally, the thermopower of different multi-terminal setups subject to magnetic fields has been measured~\cite{maximov_low-field_2004,goswami_signatures_2011,matthews_experimental_2013}.

Here, we establish a quantum analogue of the classical Nernst engine based on a four-terminal cross structure in a two-dimensional electron gas in the quantum Hall regime where transport occurs along dissipationless edge channels. As energy-dependent scattering is a prerequisite for any thermoelectric response in order to break particle-hole symmetry, we analyze different ways of introducing it into the system. The central questions we address are what asymmetry the quantum Nernst engine exhibits and what bounds for the maximal efficiency and \EMP derive from it. We also check if the different realizations of quantum Nernst engines that we consider saturate these bounds.

\section{\label{sec:Nernst}Mesoscopic Nernst effect}
We start by briefly reviewing the scattering theory of thermoelectric transport~\cite{buttiker_four-terminal_1986,butcher_thermal_1990}. We consider a central scattering region connected to single-channel electric leads $i$ with chemical potential $\mu+eV_i$ and temperature $T+\Delta T_i$. In linear response in the applied voltage and temperature bias, the charge and heat current, $I^{e,h}_i$, can be compactly written as
\begin{align}
	\label{eq:pcurrent}
	I^e_i=\sum_j e\left(L^{eV}_{ij}F^V_j+L^{eT}_{ij}F^T_j\right),\\
	\label{eq:hcurrent}
	I^h_i=\sum_j \left(L^{hV}_{ij}F^V_j+L^{hT}_{ij}F^T_j\right).
\end{align}
In Eq.~\eqref{eq:pcurrent} and~\eqref{eq:hcurrent} 
$F^V_i=eV_i/\kBT$ and $F^T_i=\kB\Delta T_i/(\kBT)^2$ denote the affinities that drive particle and heat currents. Furthermore, we introduced the linear-response Onsager coefficients
\begin{equation}\label{eq:Onsager}
	\left(
	\begin{array}{cc}
		L^{eV}_{ij} & L^{eT}_{ij} \\
		L^{hV}_{ij} & L^{hT}_{ij}
	\end{array}
	\right)
	=
	\frac{1}{h}\int dE
	\left(
	\begin{array}{cc}
		1 & E \\
		E & E^2
	\end{array}
	\right)
	\frac{\delta_{ij}-T_{i\leftarrow j}(E)}{4\cosh^2\frac{E}{2\kBT}},
\end{equation}
that link currents to affinities. Here, $T_{i\leftarrow j}(E)$ denotes the energy-dependent transmission probability from lead $j$ to lead $i$.
We remark that within linear response heat and energy currents are identical.

\begin{figure}
	\includegraphics[width=\columnwidth]{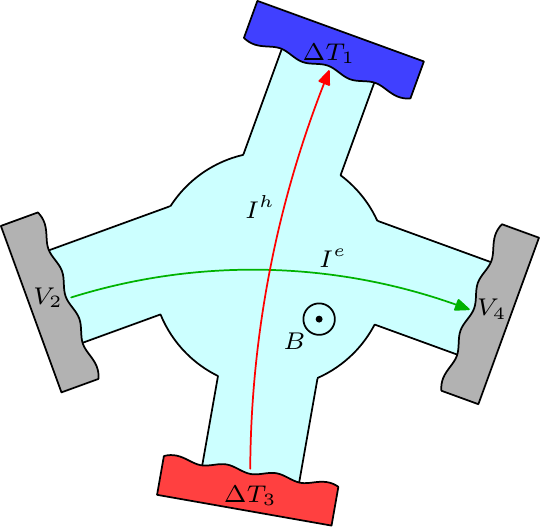}
	\caption{\label{fig:model}Schematic sketch of a generic mesoscopic Nernst engine. Four terminals are connected to a scattering region subject to a perpendicular magnetic field. A temperature bias is applied between terminals 1 and 3 while a voltage bias is applied between terminals 2 and 4. A heat current flows between terminals 1 and 3. In addition, a charge current flows between terminals 2 and 4.}
\end{figure}
We now consider more specifically the case of a mesoscopic Nernst engine. It consists of a four-terminal structure, cf. Fig~\ref{fig:model}, with a temperature bias appplied between contacts 1 and 3. In addition, a bias voltage can be applied between ccontacts 2 and 4 to generate a finite output power. For a Nernst engine, we impose the boundary conditions $I^e_1=I^e_3=0$, i.e., terminals 1 and 3 act as voltage probes~\cite{buttiker_four-terminal_1986}. At the same time, we have the boundary conditions $I^h_2=I^h_4=0$, i.e., terminals 2 and 4 act as temperature probes~\cite{engquist_definition_1981}.

Eliminating $F^V_1$, $F^V_3$, $F^T_2$, $F^T_4$ using the boundary conditions and setting the charge current $I^e=I^e_4$ and heat current $I^h=I^h_3$, we obtain
\begin{equation}
	\mathbf I=\boldsymbol{\mathcal L} \mathbf F
\end{equation}
with $\mathbf I=(I^e,I^h)^T$, $\mathbf F=(F^V_4-F^V_2,F^T_3-F^T_1)^T$ and
\begin{equation}
	\boldsymbol{\mathcal L}=\left(
	\begin{array}{cc}
		\mathcal L^{eV} & \mathcal L^{eT} \\
		\mathcal L^{hV} & \mathcal L^{hT}
	\end{array}
	\right).
\end{equation}
The Onsager coefficients $\mathcal L$ can be obtained from the coefficients $L$ given in Eq.~\ref{eq:Onsager}. We discuss their explicit forms in the specific examples discussed below. Due to the presence of a magnetic field and inelastic scattering at the probe terminals, we have $\mathcal L^{eT}\neq e\mathcal L^{hV}$ at a given magnetic field. This is permitted by the Onsager relations which only relate the off-diagonal coefficients to one another for a reversed magnetic field~\cite{casimir_onsagers_1945}. In the following, we discuss the implications of this property for the maximal power, maximal efficiency and \EMP.

The power the heat-driven current delivers by performing work against a bias voltage $F^V_4-F^V_2$ is given by
\begin{equation}
	P=-I^e\frac{\kBT}{e}(F^V_4-F^V_2).
\end{equation}
The corresponding efficiency of heat-to-work conversion is given by the ratio between output power and input heat, $\eta=P/I^h$. As was shown in Ref.~\cite{benenti_thermodynamic_2011}, the maximal efficiency is given by
\begin{equation}
	\eta_\text{max}=\eta_Cx\frac{\sqrt{1+y}-1}{\sqrt{1+y}+1},
\end{equation}
where $y=\mathcal L^{eT}\mathcal L^{hV}/\Det \boldsymbol{\mathcal L}$ is the generalized figure of merit and $x=\mathcal L^{eT}/(e\mathcal L^{hV})$ denotes the asymmetry of the off-diagonal Onsager coefficients. Optimizing the output power with respect to the applied bias for a fixed temperature difference, we get the maximal power
\begin{equation}
	P_\text{max}=\frac{1}{4}\frac{\kBT}{e}\frac{(\mathcal L^{eT})^2}{\mathcal L^{eV}}(F^T_3-F^T_1)^2,
\end{equation}
and the associated \EMP
\begin{equation}
	\eta_\text{maxP}=\frac{\eta_C}{2}\frac{xy}{2+y}.
\end{equation}

The second law of thermodynamics requires the rate of entropy production to be non-negative, $\dot S=\mathbf F^T\mathbf I=\mathbf F^T\boldsymbol{\mathcal L}\mathbf F\geq0$. This imposes the condition that $\mathcal L$ is positive semidefinite. This translates into the condition $0\leq y \leq h(x)$ with $h(x)=4x/(1-x)^2$ for $x>0$ and $h(x)\leq y\leq 0$ for $x<0$. Hence, for $|x|>1$, the \EMP can be larger than $\eta_C/2$ and even reach Carnot efficiency in the limit $x\to\infty$. However, as was shown in Ref.~\cite{stark_classical_2014}, for a Nernst engine the conservation of current imposes the additional constraint that the matrix $\boldsymbol{\mathcal K}_1=\boldsymbol{\mathcal L}+\boldsymbol{\mathcal L}^T+i(\boldsymbol{\mathcal L}-\boldsymbol{\mathcal L}^T)$ is positive semi-definite. This gives rise to the condition 
$y\leq2x/(1-x)^2$ and therefore $\eta_\text{max}\leq\eta_C\left(1-x+x^2-|1-x|\sqrt{1+x^2}\right)$ as well as $\eta_\text{maxP}\leq (\eta_C/2) x^2/(x^2-x+1)$.

For some of the systems discussed below we will find that instead of the matrix $\boldsymbol{\mathcal K}_1$ being positive semi-definite, they fulfill the stronger condition that the matrix $\boldsymbol{\mathcal K}_\alpha=\boldsymbol{\mathcal L}+\boldsymbol{\mathcal L}^T+i\sqrt{\alpha}(\boldsymbol{\mathcal L}-\boldsymbol{\mathcal L}^T)$ with some parameter $\alpha\geq1$ is positive semi-definite as well. This leads to the slightly more general bound $y\leq4x/[(1+\alpha)(1-x)^2]$ and, hence,
\begin{multline}
	\eta_\text{max}\leq\frac{\eta_C}{2}\left[1+x^2+\alpha(1-x)^2-\right.\\\left.
	|1-x|\sqrt{1+\alpha}\sqrt{(1+x)^2+\alpha(1-x)^2}\right],
\end{multline}
as well as
\begin{equation}
	\eta_\text{maxP}\leq\frac{\eta_C}{2}\frac{2x^2}{1+x^2+\alpha(1-x)^2}.
\end{equation}

\section{\label{sec:Results}Results}
In the following we analyze the thermoelectric performance of different types of quantum Nernst engines. We will start by a setup where two of the arms of a cross structure contain quantum point contacts. We then discuss the case in which an antidot is embedded into the middle of the cross structure.

\subsection{\label{ssec:QPC}Quantum point contacts}
\begin{figure}
	\centering
	\includegraphics[width=.85\columnwidth]{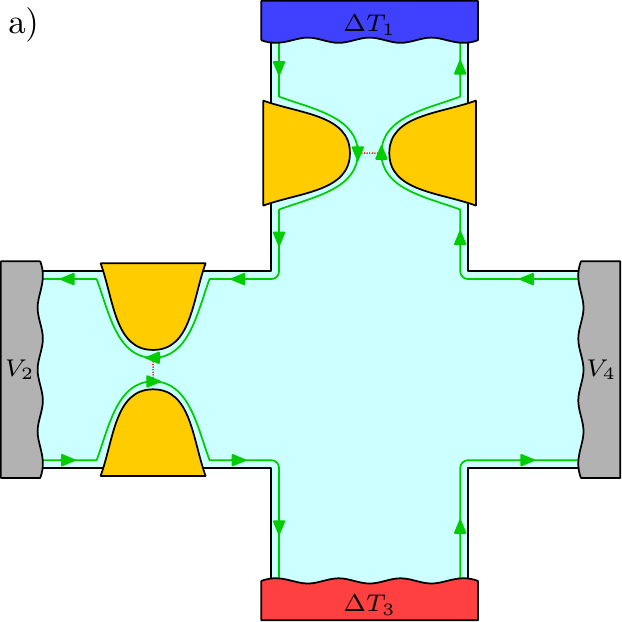}
	\includegraphics[width=\columnwidth]{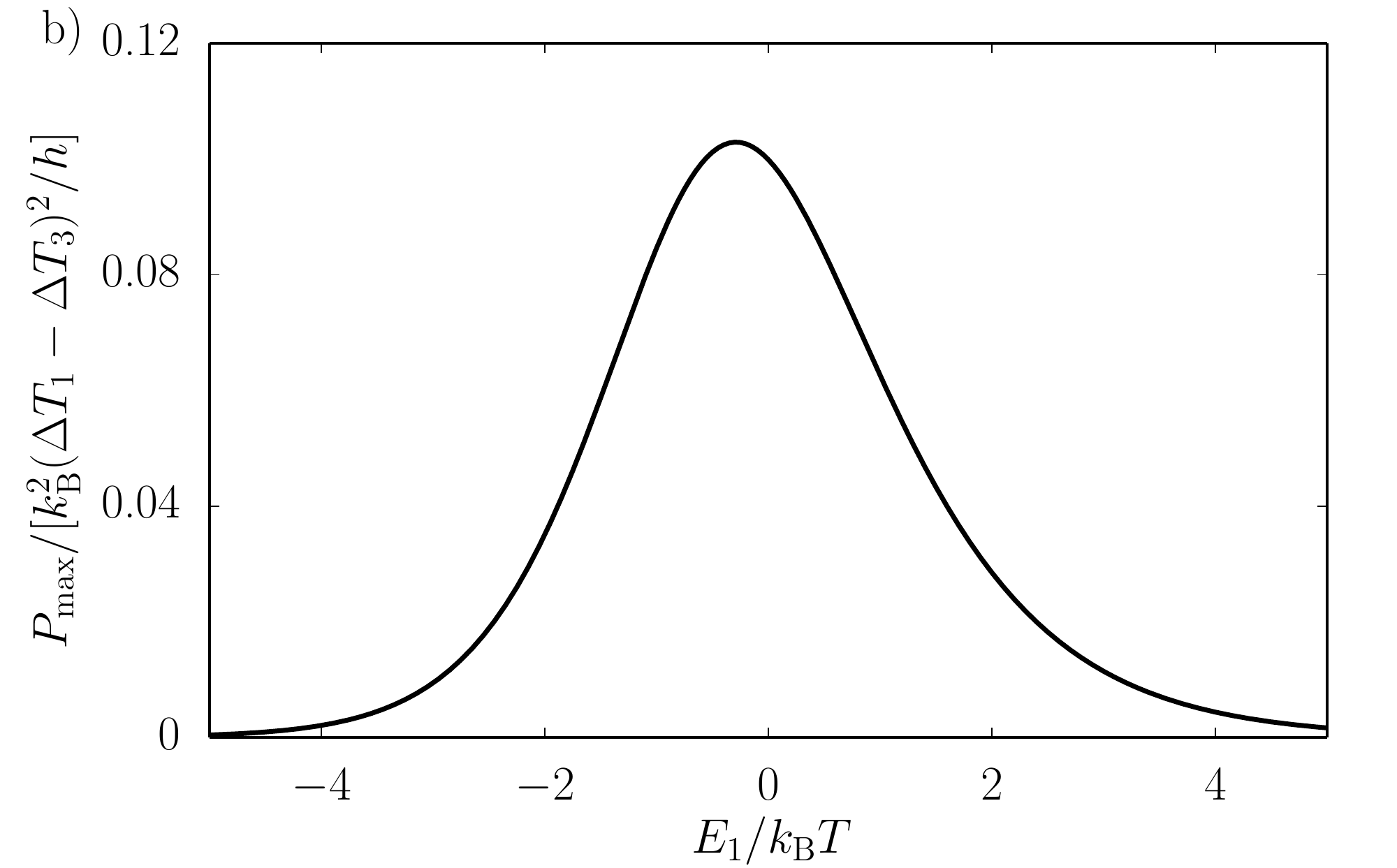}
	\includegraphics[width=\columnwidth]{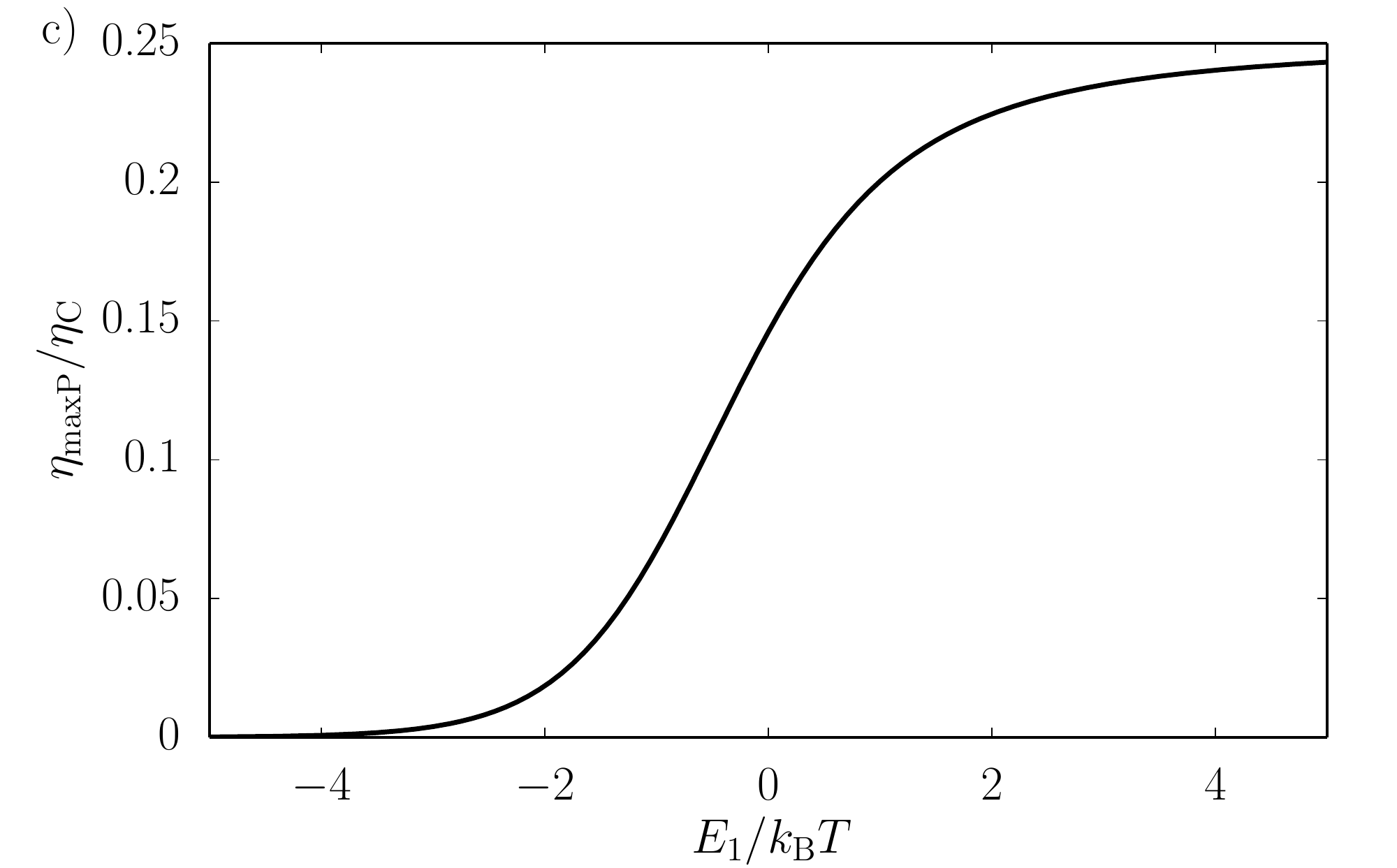}
	\caption{\label{fig:Nernst1}(a) Quantum Nernst engine I. Edge channels in a 2-dimensional electron gas in the quantum Hall regime propagate in a four-terminal cross geometry with two quantum point contacts. Terminals 1 and 3 are temperature-biased and inject heat but no charge current while terminals 2 and 4 are voltage-biased and inject charge but no heat current into the sample. (b) Maximal power and (c) EMP as a function of the threshold energy $E_1$ for $E_2=E_1$.}
\end{figure}
The first quantum Nernst engine we consider has a quantum point contact embedded into arms 1 and 2 of the cross structure, cf. Fig.~\ref{fig:Nernst1}(a). For simplicity, we model the energy dependent transmission of the quantum point contacts by a step function with threshold energy $E_1$ and $E_2$, respectively. We find qualitatively similar results in the case of a more realistic transmission, such as the saddle point potential model of a quantum point contact~\cite{buttiker_quantized_1990}. The Onsager coefficients of the Nernst engine for $E_1>E_2$ are found to be (for $E_2>E_1$ the roles of $E_1$ and $E_2$ in the expression for $\mathcal L^{eT}$ are simply exchanged)
\begin{equation}
	\boldsymbol{\mathcal L}_\text{qpc}=\left(
	\begin{array}{cc}
		\frac{e}{h}\frac{g_{1,E_2}g_{3,E_2}-g_{2,E_2}^2}{g_{3,E_2}} & \frac{e}{h}\frac{g_{2,E_2}(g_{1,E_1}g_{3,E_1}-g_{2,E_1}^2)}{g_{1,E_1}g_{3,E_2}}\\
		0 & \frac{1}{h}\frac{g_{1,E_1}g_{3,E_1}-g_{2,E_1}^2}{g_{1,E_1}}
	\end{array}
	\right),
\end{equation}
where the analytic expressions for the functions $g_{n,y}$ are given in the Appendix.

In contrast to the classical Nernst engine which has $\mathcal L^{eT}=-e\mathcal L^{hV}$~\cite{stark_classical_2014}, for the quantum Nernst engine we find $\mathcal L^{hV}=0$ for our choice of magnetic field direction. Hence, the asymmetry parameter $x$ diverges, $x\to\infty$. As a consequence, the maximal efficiency and the \EMP coincide. As discussed above, current conservation implies that the matrix $\boldsymbol{\mathcal K}_1$ is positive semidefinite and, hence, imposes the upper bounds $\eta_\text{max}=\eta_\text{maxP}\leq\eta_C/2$. However, for the specific form of the Onsager coefficients $\boldsymbol{\mathcal L}$ we find that even the matrix $\boldsymbol{\mathcal K}_3$ is positive semidefinite and hence the stronger bounds $\eta_\text{max}=\eta_\text{maxP}\leq\eta_C/4$ apply.

We now want to address the question if the efficiency bounds that we have just established are saturated. To this end, we show in Fig.~\ref{fig:Nernst1}(b) and (c) the maximal power and \EMP as a function of the threshold energy of QPC 1 for $E_1=E_2$ (we checked that both the output power and the efficiency are maximal along this line). We find that the system delivers a maximal power of  $P_\text{max}\approx0.1(\kB\Delta T_1-\kB\Delta T_3)^2/h$ for $E_1\approx -0.3\kBT$. At this point, the device has an efficiency of $\eta_\text{maxP}\approx0.12\eta_C$. For $E_1=E_2\to\infty$ the largest \EMP of $\eta_\text{max}=\eta_C/4$ is reached, i.e. the bound deriving from $\boldsymbol{\mathcal K}_3$ being positive-semidefinite is indeed saturated. Interestingly, the quantum Nernst engine we propose here outperforms the classical Nernst engine discussed in Ref.~\cite{stark_classical_2014} in terms of efficiency even if it does not saturate the bounds that derive from current conservation alone. We remark that similarly to the classical Nernst engine the output power in the regime of largest \EMP is exponentially suppressed.

We finally briefly comment on the fluctuation-disspation theorem in our setup. From scattering theory, we derive that the equilibrium correlations between charge and heat currents satisfy~\cite{blanter_shot_2000}
\begin{align}
	S_{I^eI^e}^\text{eq}&=4e\mathcal L^{eV},\\
	S_{I^hI^h}^\text{eq}&=4\kBT\mathcal L^{hT},\\
	S_{I^eI^h}^\text{eq}=S_{I^hI^e}^\text{eq}&=2\kBT(\mathcal L^{eT}+e\mathcal L^{hV}).
\end{align}
This means that for the cross correlations we have to use the symmetrized Onsager coefficients. In our setup one of the off-diagonal Onsager coefficiencts vanishes. This implies that the cross correlations between heat and charge are only half as big as one could expect naively.

\subsection{\label{ssec:Antidot}Antidot}
\begin{figure}
	\centering
	\includegraphics[width=.75\columnwidth]{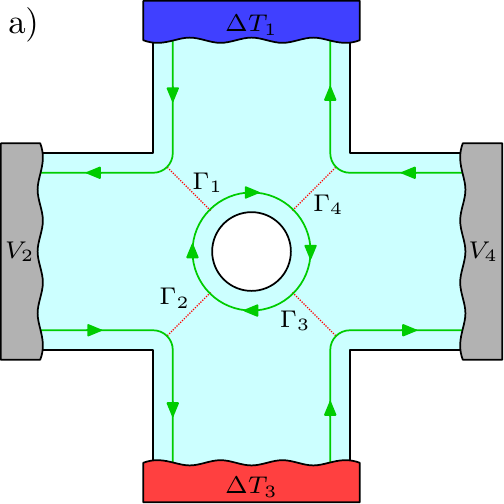}
	\includegraphics[width=\columnwidth]{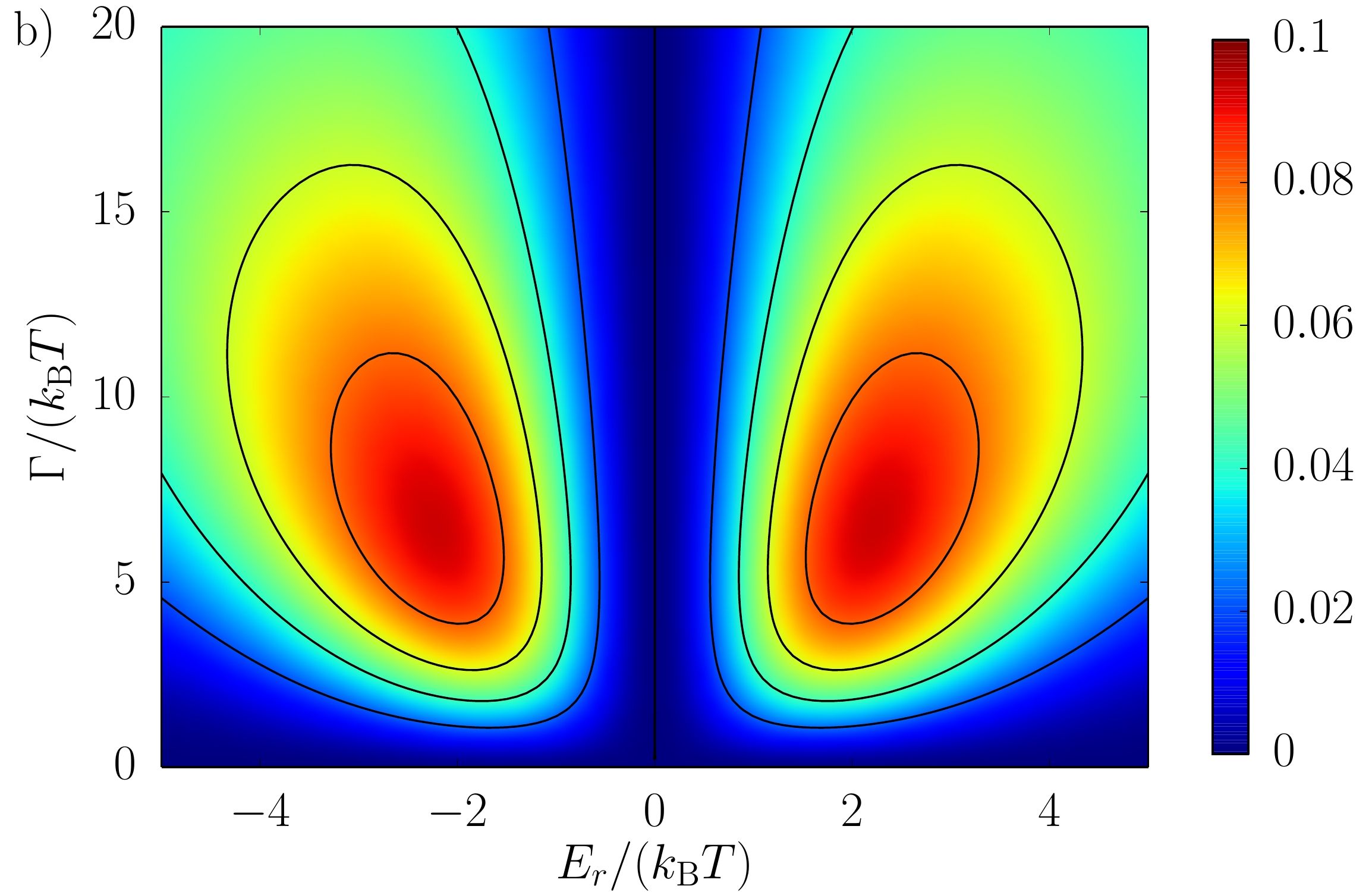}
	\includegraphics[width=\columnwidth]{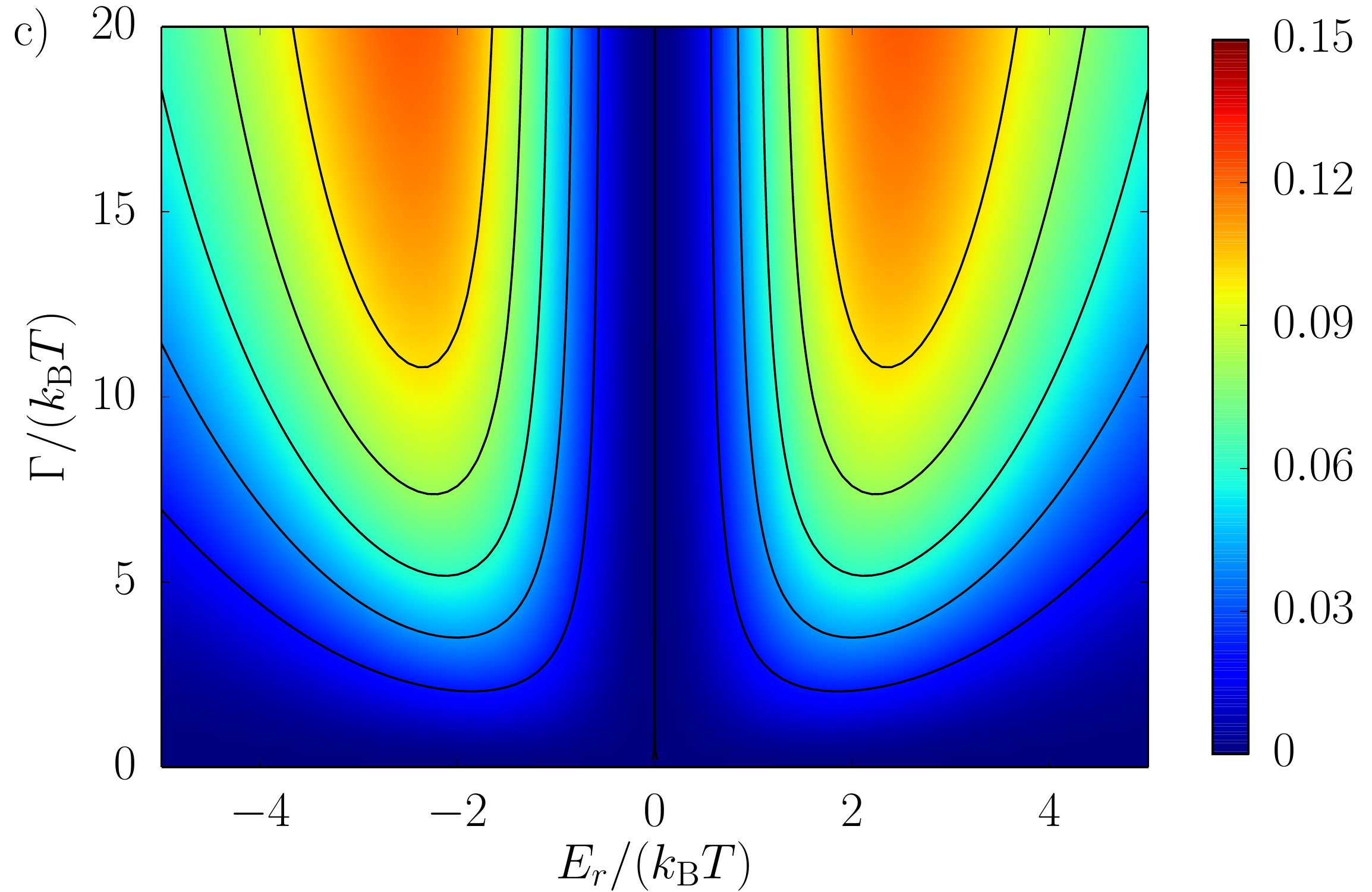}
	\caption{\label{fig:Nernst2}(a) Quantum Nernst engine II. The edge states are tunnel-coupled to an antidot formed in the middle of the cross. (b) Maximum power in units of $\kB^2(\Delta T_1-\Delta T_3)^2/h$ and (c) EMP in units of $\eta_C$ as a function of the antidot resonance position and width for $\Gamma_1=\Gamma_3=\Gamma/2$.}
\end{figure}
We now turn to second type of Nernst engine which demonstrates that for systems with broken time-reversal symmetry better energy filtering does not necessarily lead to higher efficiency~\cite{benenti_thermodynamic_2011}. This is in direct contrast to time-reversal symmetric systems where stronger energy filtering leads to higher efficiency. The setup consists of a quantum Hall cross structure with an antidot embedded in the middle, cf. Fig.~\ref{fig:Nernst2}(a). 
A related setup with helical edge channels in a two-dimensional topological insulator and different boundary conditions has been recently discussed in Ref.~\cite{hwang_nonlinear_2014}.
The antidot has a resonance with energy $E_r$ which is coupled to the edge states with coupling strength $\Gamma_i$. For $j\neq i+1$ we have $T_{j\leftarrow i}(E)=\Gamma_i\Gamma_{j-1}/\Delta$ where $\Delta=(E-E_r)^2+\Gamma^2/4$ with $\Gamma=\sum_i\Gamma_i$ while for $j=i+1$ we have $T_{i+1\leftarrow i}(E)=1-\Gamma_i(\Gamma-\Gamma_i)/\Delta$ where all indices are to be taken modulo 4.~\cite{buttiker_negative_1988}

The general expressions for $\boldsymbol{\mathcal L}$ are rather lengthy. We can obtain compact expressions in the limit $\Gamma_2=\Gamma_4=0$ which we numerically found to yield the largest possible output power and efficiency. In this limit, we have
\begin{equation}
	\boldsymbol{\mathcal L}_\text{dot}=\frac{1}{h}\left(
	\begin{array}{cc}
		e\left(f_1-\Gamma_1\Gamma_3h_1\right) & e\Gamma_1\Gamma_3h_2 \\
		\Gamma_1\Gamma_3h_2 & f_3-\Gamma_1\Gamma_3h_3
	\end{array}
	\right),
\end{equation}
with the functions $f_n$ and $h_n$ defined in the Appendix. As we have $\mathcal L^{eT}=e\mathcal L^{hV}$, the asymmetry parameter is $x=1$ and hence, the upper bound for the \EMP is $\eta_C/2$ as in the case of time-reversal symmetric systems. In the following, we are going to address whether this upper bound can be saturated when varying the position and width of the antidot resonance.

Fig.~\ref{fig:Nernst2}(b) shows the maximum power as a function of the resonance position and width. For a given width, the output power vanishes at the particle-hole symmetric point $E_r=0$ and takes maximal value at $E_r\approx\pm2\kBT$. Starting from small resonance widths, this maximal value increases until it reaches its largest value for about $\Gamma\approx 7\kBT$ and then decreases again for larger values of the resonance width.

The \EMP (which for this system we numerically found to be close to the maximal efficiency) shown in Fig.~\ref{fig:Nernst2}(c) similarly vanishes at $E_r=0$ and takes maximal values for $E_r\approx \pm2.5\kBT$ for a given resonance width. It is significantly smaller than the bounds derived from current conservation. Surprisingly, the efficiency of the system monotonically grows with the resonance width. Hence, in contrast to typical heat engines, for this Nernst engine, better energy filtering does \emph{not} lead to higher efficiency. 

We now elucidate the mechanism behind this seemingly paradoxical result. For a broad resonance, most electrons are transmitted from terminal 3 (4) to 2 (1) and vice versa. Due to the boundary conditions $I^p_1=I^p_3=I^h_2=I^h_4=0$, these processes do not contribute either to the heat or charge current in the system. Only processes where electrons follow a different path through the system can give rise to any thermoelectric response. Due to the broad resonance, these processes most likely involve high-energy electrons. The number of electrons with a given energy decays exponentially with this energy. Hence, effectively only electrons in a certain energy window contribute to the thermoelectric response, thereby leading to a high efficiency.

We finally mention that our antidot heat engine does not saturate the general bounds on the \EMP that follow from thermodynamics. This is not surprising since for a system with $x=1$ the limit $\eta_\text{maxP}=\eta_C/2$ is only reached when heat and charge currents are proportional to each other~\cite{van_den_broeck_thermodynamic_2005}, due to, e.g., transport through a sharp level, which is clearly not the case in our system.

\section{\label{sec:Conclusions}Conclusions}
We analyzed a quantum Nernst engine based on thermoelectric transport along quantum Hall edge states. We first considered a setup based on quantum point contacts which we found to outperform a recently proposed classical Nernst engine in terms of efficiency.~\cite{stark_classical_2014} It exhibits an extreme asymmetry of the off-diagonal Onsager coefficients. For a second setup based on an antidot, we demonstrated that a better energy filtering does not necessarily lead to a larger efficiency for systems with broken time-reversal symmetry.

\acknowledgments
We acknowledge interesting disussions with K. Brandner as well as financial support from the Swiss NSF via the NCCR QSIT, the Spanish MICINN Juan de la Cierva program and MAT2011-24331, the COST action MP1209 and the US NSF grant DMR-0844899.

\renewcommand{\thetable}{A\arabic{table}}
\section{Appendix: Integrals}
The functions $f_n$, $g_{n,y}$ and $h_n$ used in the main text are defined in terms of the following integrals,
\begin{align}
	f_n&=\int_{-\infty}^\infty dx \frac{x^{n-1}}{4\cosh^2\frac{x}{2\kBT}},\\
	g_{n,y}&=\int_y^\infty dx \frac{x^{n-1}}{4\cosh^2\frac{x}{2\kBT}},\\
	h_n&=\int_{-\infty}^\infty dx \frac{1}{(x-E_r)^2+\Gamma^2/4}\frac{x^{n-1}}{4\cosh^2\frac{x}{2\kBT}}.
\end{align}
For the functions $f_n$ and $g_{n,y}$, the following compact analytical expressions can be obtained for $n=1,2,3$:
\begin{align}
	f_1&=\kBT,\\
	f_2&=0,\\
	f_3&=\frac{\pi^2}{3}(\kBT)^3,\\
	g_{1,y}&=\frac{\kBT}{1+e^{y/(\kBT)}},
	\\
	g_{2,y}&=\frac{(\kBT)^2}{2}\left(\log 4+2\log\cosh\frac{y}{2\kBT}\right.\\
	&\phantom{=}\left.-\frac{y}{\kBT}\tanh\frac{y}{2\kBT}\right),
	\\
	g_{3,y}&=\frac{(\kBT)^3}{2}\left\{-4\Li_2\left(-e^{-y/(\kBT)}\right)+\frac{y}{\kBT}\right.\\\phantom{=}&\left.
	\left[\frac{y}{\kBT}+4\log\left(1+e^{-y/(\kBT)}\right)-\frac{y}{\kBT}\tanh\frac{y}{2\kBT}\right]\right\}.
\end{align}


\end{document}